\documentclass[12pt]{iopart}
\usepackage{graphicx}
\begin{document}

\title{critRHIC: The RHIC Low Energy Program}

\author{G S F Stephans}

\address{Laboratory for Nuclear Science, Massachusetts Institute of Technology, Cambridge, MA 02139 USA}
\ead{gsfs@mit.edu}
\begin{abstract}

  Recent experimental and theoretical developments have motivated interest in
  a more detailed exploration of heavy ion collisions in the range
  $\sqrt{s_{_{NN}}}$=5--15~GeV.  In contrast to interactions at the full RHIC
  energy of $\sqrt{s_{_{NN}}}$=200~GeV, such collisions result in systems
  characterized by much higher baryon chemical potential, $\mu_B$.  Extensions
  of lattice QCD calculations to non-zero values of $\mu_B$ suggest that a
  critical point may exist in this region of the QCD phase diagram.  Discovery
  of the critical point or, equivalently, determining the location where the
  phase transition from partonic to hadronic matter switches from a smooth
  crossover to 1$^{\rm st}$ order would establish a major landmark in the phase
  diagram.  Initial studies of Pb+Pb collisions in this energy range have
  revealed several unexpected features in the data.  In response to these
  results, it has been suggested that the existing RHIC accelerator and
  experiments can be used to further the investigation of this important
  physics topic.  This proceeding briefly summarizes the theoretical and
  experimental situation with particular emphasis on the conclusions from a
  RIKEN BNL workshop held in March of 2006.

\end{abstract}

\submitto{\jpg}

\section{Introduction}
The primary goal of experimental studies of heavy ion collisions at relativistic
energies is the exploration of QCD matter at extremes of high temperature and
density, in effect to map out a section of the QCD phase diagram.  Recent work
at the Relativistic Heavy Ion Collider (RHIC) at Brookhaven National
Laboratory (BNL) has explored the high temperature, low baryon density regime
\cite{WhitePaper, WhiteBNL}.  The data show evidence for the creation of a
novel equilibrated medium whose primary degrees of freedom are partonic.  The
constituents of this matter interact unexpectedly strongly with each other and
it is almost opaque to high energy partons.  Work exploring the properties of
this new state of matter and the conditions necessary for its creation are
ongoing.  Simultaneously, recent experimental and theoretical developments
have suggested that equally important discoveries are possible at higher
baryon chemical potential, $\mu_B$.  Specifically, lattice QCD calculations
suggest the presence of a critical point where the phase transition from
partonic to hadronic degrees of freedom switches from the smooth crossover
believed to exist at RHIC to a 1$^{\rm st}$ order transition.  On the experimental
side, a variety of data in the range $\sqrt{s_{_{NN}}}$=5--15~GeV, primarily
unexpectedly non-monotonic behavior of several observables, hint at the
presence of some novel processes.  These factors led to the suggestion that
the RHIC program might be extended to explore this additional region of the
QCD phase diagram \cite{GSFS_Fixed}.  A detailed discussion of both the
experimental and theoretical issues took place at a RIKEN BNL Research Center
Workshop held from March 9--10, 2006 \cite{BNLWorkshop}.  The workshop program
is given in Table~\ref{Program}.

\section{Theory of the QCD Critical Point}
The possibility of a critical point in the QCD phase diagram is not a recent
idea.  While not specifically related to the critical point of current
interest, early work used the so-called Glasgow method to extend QCD studies
to non-zero $\mu_B$ \cite{GlasRev}.  Other more directly related theoretical
work is reviewed in \cite{StephanovRev}.  It was also realized early on that
heavy ion collisions could be used to explore this physics \cite{critHI}.
Later, efforts began to extend lattice QCD into the high baryon density region
\cite{FodKat01}, and this remains an area of active interest
\cite{Allton,FodKat04,GG05,Redlich, MAS, Fujii, Philip}.  The existence of the critical point
itself is not generally questioned since the presence of a smooth cross-over
at $\mu_B\approx0$ is well established from lattice QCD calculations while
there are equally valid, although less rigorous, expectations of a 1$^{\rm
  st}$ order transition at very high $\mu_b$ and very low temperature
\cite{highMu}.  Therefore, efforts are focused on determining the location and
characteristics of the critical point.  This remains very much a work in
progress but values in the vicinity of $\mu_B \approx 450 \pm 250$~MeV appear
quite reasonable.  This is exactly the range at chemical freezeout extracted
from fitting ratios of particles emitted in heavy ion collisions at
$\sqrt{s_{_{NN}}}$=5--15~GeV \cite{Thermal}.  Preliminary indications are that
the temperature of the critical point is not too far above those extracted
from particle ratio data.  Provided that the critical point and its associated
1$^{\rm st}$ order phase transition line are not located at very high $\mu_B$
or at temperatures far above the chemical freezeout point, the systems created
in heavy ion collisions should be sensitive to their influence.

\section{Existing Data at High $\mu_B$}
Another motivation for the interest in exploration of the high $\mu_B$ region
of the phase diagram arises from unexpected features in several experimental
observables measured by the NA49 experiment at the SPS \cite{NA49SQM}.  Data
were taken for Pb+Pb using a fixed target and beam energies of 20, 30, 40, 80,
and 158 A$\cdot$GeV.  The results showed significant deviations from
monotonicity, especially in the spectral shapes for kaons and the ratios of
kaons to pions, which are difficult to reproduce in existing models
\cite{SandSQM}.  A preliminary extraction of event-by-event fluctuations in
the K$^+$ over $\pi^+$ ratio showed a significant excess over the expectations
from hadronic cascade models \cite{KpiFluct}.  Since these data were
covered by other talks at this conference, they will not be
discussed further.

\section{Experimental Signatures of the Phase Structure}
At present, there are many suggestions for how to search experimentally for
evidence of the critical point and/or the 1$^{\rm st}$ order phase transition.  Most of
the ideas are more qualitative in nature.  Although the manifestations of the
phase structure are unambiguous in some models, making quantitative connections
to specific observables remains challenging.  Not surprisingly, many of the
proposals relate to fluctuations and correlations, including particle ratios,
multiplicity, baryon number, and transverse momentum\cite{Fl1, Fl2, Fl3, Fl4, Fl5}.  In addition, bulk
properties such as directed and elliptic flow, especially comparing pions to
protons, are expected to be sensitive to this physics \cite{Stocker, Shuryak}.  One
aspect of particular interest to the experimental design is that the
properties of initial interest are predominantly more global in nature and
therefore do not require large event samples.  Two aspects of the critical
point itself may have significant consequences.  First, although signals such
as large susceptibilities peak exactly at the point, their values are significantly
different from the average over an extended region in both temperature and
$\mu_B$ \cite{HattaIkeda}.  Second, hydrodynamic calculations suggest that the
phase space trajectories of evolving hot and dense QCD matter may be attracted to the
vicinity of the critical point \cite{Nonaka}.  If true, this implies that an
experimental scan will not need to use very fine steps in $\sqrt{s_{_{NN}}}$
to find evidence of the phase structure.

\section{Machine and Detector Capabilities}
By far the most important consideration in planning a low energy physics
program at RHIC concerns the capability of the Tandem/AGS/RHIC complex to
accelerate and collide ions at almost 2 orders of magnitude below the full
RHIC design energy.  From the start, the accelerator physicists focused their
studies on the idea of injecting and colliding at each separate beam energy in RHIC,
rather than injecting at a single energy and accelerating as was done for the
higher energy running.  The Tandem/AGS combination had previously been used to
generate Au beams down to 2~A$\cdot$GeV for the fixed target program
so the only questions concerned transferring, circulating, and colliding the
ions in RHIC.  The very low beam rigidities present complications of magnet
and power supply stability at low current and the low velocities result in
beam physics issues of emittance and intrabeam scattering.

\begin{figure}[htb]
\begin{center}
\includegraphics[width=2.5in]{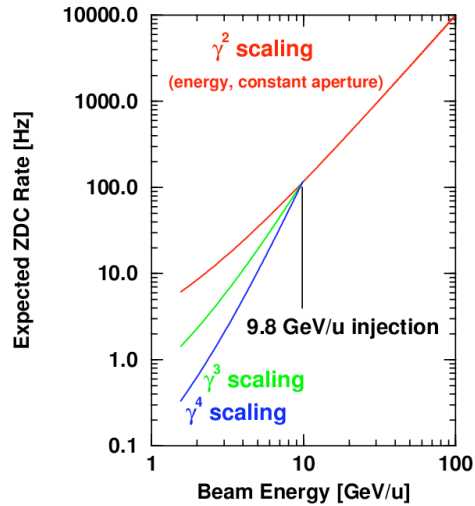}
\caption{Projections of minimum bias Au+Au event rate as a function of
 the energy in one beam (i.e. half of $\sqrt{s_{_{NN}}}$).  The line at the upper right represents rates already
  achieved, which scale closely with $\gamma^2$.  At lower energies, the rate
  is expected to drop more quickly.  The lines shown represent scaling with
  $\gamma^3$ and  $\gamma^4$.}
\label{BeamRate}
\end{center}
\end{figure}

These potential problems have been evaluated and no clear obstacles to running
RHIC in the range $\sqrt{s_{_{NN}}}$=5--15~GeV were found.  The emittance of
the beam out of the AGS at the lowest energy is large, typically
5--6~$\pi$mm~mrad, but this does not exceed the acceptance of the transfer
line or the injection system for RHIC.  Although some work will likely be
needed on improving power supply quality, this is a well understood process.
Tests described below indicate that there are no dramatic unexpected magnet
problems.  One disappointment in the early studies was that injecting Au ions
which were not fully stripped will not be effective.  This was investigated as
a way to allow higher magnet currents at the lowest energies but calculations
indicate that the RHIC vacuum is not good enough for stable circulating beams
with remaining electrons.  For fully stripped ions, beam lifetimes are
expected to be more than sufficient for a physics program, with the dominant
luminosity loss mechanism being the expansion of the bunches due to space
charge effects.  One challenge which remains to be solved is luminosity
monitoring, which is done at higher energies using the Zero Degree
Calorimeters.  The resolution of the ZDC drops with beam energy and the
increasing spread of the neutrons due to Fermi motion causes significant
inefficiencies due to the limited aperture of the detector.

The most difficult aspects to evaluate without testing involve intrabeam
scattering, bunch stability, beam profile, and other effects which impact the
luminosity.  The expected range is illustrated in Fig.~\ref{BeamRate} which
shows the minimum bias Au+Au event rate as a function of the energy in an
individual beam (i.e. half of $\sqrt{s_{_{NN}}}$).  The line starting at the
upper right shows rates already achieved in the range from the normal
injection energy of 9.8~A$\cdot$GeV up to full RHIC energy.  Recent
improvements should allow raising these values by factors of $\sim$2--5.  The
observed scaling with $\gamma^2$ is well understood and dominated by
emittance and the fixed machine aperture.  Below the current injection
energy, it is expected that intrabeam scattering and other effects may be more
dominant so that the rate will drop more quickly.  Two possibilities (scaling
with $\gamma^3$ and  $\gamma^4$) are shown for illustration.  Studies with
colliding ion beams will be needed before the achievable rates can be better
established.

One distinct advantage of lower energy beams is that the luminosity can be
improved much more easily using electron cooling.  Prototyping and testing is
underway at BNL to develop a cooling system to increase the full energy
luminosity for the future RHIC-II and eRHIC facilities.  Cooling of lower
energy ion beams is easier technologically and therefore much simpler and less
expensive to implement.  Potential exists to increase low energy luminosity
using cooling in both the AGS and RHIC.  It is possible that existing
equipment, either from other machines or being built as part of the RHIC-II
development project, could provide most of the necessary hardware.
Simulations have been performed which show that a fairly small installation
could achieve dramatic reductions in the emittance in times of the order of
minutes or less.

The overall conclusion of the machine evaluation to date is very encouraging.
No major difficulties which would prevent running RHIC in collider mode at
significantly lower energies were found.  The projected rates are more than
sufficient for an initial physics program and several possibilities of future
improvements are available.

The RHIC program has been a spectacular success with the detectors producing a
wealth of high quality physics results from a plethora of colliding systems.
Since the physics observables and analyses of interest for lower energy
running are largely those already studied at RHIC, the capability to perform
the required measurements is well established.  Of particular note is that
many of these analyses have also been performed for p+p collisions which have
much lower multiplicities.  Therefore, even the needed adjustments for lower
numbers of particles per event have been extensively investigated.  Some
adjustments to the normal Au+Au mode of operations, especially in the areas of
triggering and centrality selection, must be made, but again the p+p experience
will help.  This area of the program appears to have
almost no uncertainties to address.

One important issue that must be considered is the relative merit of studying
the energy range of interest using colliding beams or fixed targets.  The
latter have a clear advantage in event rate which can in some cases be
trivially boosted using multiple or thicker targets.  Also, kinematic
focusing concentrates a larger fraction of the center of mass frame solid
angle into a specific detector area so larger acceptance can be achieved more
easily.  However, for measurements which do not depend on the highest possible
event rate, and in particular when scanning a range of beam energies, the
colliding beam arrangement has major advantages.  Due to the fixed
correspondence between detector geometry and the center of mass frame for the
interaction, the acceptance, location, and absolute momentum of tracks at a
given rapidity, as well as other aspects of the measurement, are independent of beam
energy.  This unchanging environment has enormous benefits in reducing
systematic errors associated with comparing results across beam energies.

\begin{figure}[htb]
\begin{center}
\includegraphics[width=2.5in]{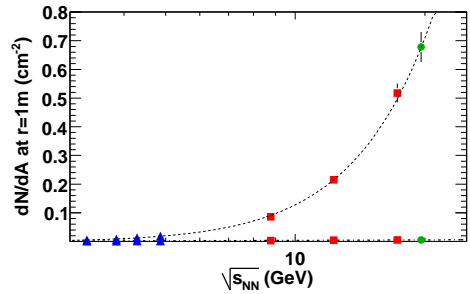}
\caption{A comparison of particle densities at a distance of 1~m from the
  interaction for fixed target (upper points) and colliding beam (lower
  points) experiments.  The beam energies range from the AGS (triangles)
  through the SPS (squares) and the lowest RHIC energy to date (circle).}
\label{HitDensity}
\end{center}
\end{figure}

An illustration of another advantage of collider geometry is shown in
Fig.~\ref{HitDensity} where the hit density on the detector is plotted versus
$\sqrt{s_{_{NN}}}$.  For colliding beams, the hit density rises very slowly
while kinematic focusing causes a much steeper dependence for fixed target
experiments.  This parameter directly relates to the detector occupancy and
average track separation which can have a significant impact on the analysis
capabilities.  In addition, troublesome detector effects such as space charge
and electronic baseline shifts, among others, get much worse as hit density
increases.  As for the effects mentioned above, this weak dependence of the
detector environment on beam energy substantially reduces the relative
systematic errors associated with an energy scan.

\section{RHIC Machine Tests}
Encouraged by the results of the RIKEN BNL workshop, RHIC management accepted
the suggestion to perform preliminary machine studies to explore the machine
capabilities for low energy running.  In early June, 2006, protons were
injected, circulated, and collided at $\sqrt{s}$=22~GeV.  The choice of beam
was dictated by the primary RHIC objective during this period which was
collisions of high luminosity polarized protons.  The magnet settings for this
test corresponded to those needed for Au+Au at about $\sqrt{s_{_{NN}}}$=9~GeV,
or equivalently fixed target running with a beam energy of $\sim$40~GeV. These
values were chosen to be significantly lower than any used previously while
sufficiently high for a reasonable chance of success.  The results of this
test were quite encouraging, with 2--3 hour lifetimes of the beam with
collisions.  No significant show-stoppers were found.  Not surprisingly, some
challenges were evident but they are ones the machine physicists know how to
approach.  In addition, extensive optics measurements were performed which
allow development of machine settings for even lower energy.  Plans for
tests at lower energy using ions rather than protons are ongoing.

\section{Related Experimental Programs}
Exploring the QCD phase diagram using low energy running at RHIC is part of a
broad heavy ion physics program.  Higher energy running at RHIC will continue in order
to extract detailed information about the novel partonic system formed at low
baryon chemical potential \cite{QM05}.  In the future, these studies will be
pushed to even higher energy using the LHC under construction at CERN
\cite{LHC}.  More directly related experimental efforts using fixed targets
are being planned both at the CERN SPS \cite{SPS} and the future CBM experiment
at the FAIR facility \cite{CMB}.

\section{Conclusions}
Recent experimental and theoretical results suggest that major discoveries are
possible in the region of high baryon density accessible using heavy ion
collisions at $\sqrt{s_{_{NN}}}$=5--15~GeV.  Evaluations of machine and
detector capabilities have shown that colliding beam experiments at RHIC have
the potential to make enormous contributions in this area.  The details of the
proposed program are under active development.

\begin{table}[ht]
\caption{RIKEN BNL Workshop: Can We Discover the QCD Critical
  Point at RHIC?}
\label{Program}
\begin{tabular}{@{}ll}
\br
Title of talk& Name\\
\mr
Can we discover the QCD critical point at RHIC? Theoretical & K.\ Rajagopal\\
Lattice results on the QCD critical point & F.\ Karsch\\
QCD critical point and correlations & M.\ Stephanov\\
Can we discover the QCD critical point at RHIC? Experimental & G.\ Roland\\
Prospects for RHIC low energy operations & T.\ Satogata\\
Energy dependence of Pb+Pb collisions at the CERN SPS & P.\ Seyboth\\
Four special points of the high-mu RHIC & E.\ Shuryak\\
Soft mode of the QCD critical point & H.\ Fujii\\
Baryon number fluctuation near the critical point & Y.\ Hatta\\
Hydrodynamical evolution near the QCD critical point & C.\ Nonaka\\
Prospects of a new ion program at the CERN SPS & M.\ Gazdzicki\\
PHENIX capabilities for the low-energy RHIC run & P.\ Steinberg\\
Search for the critical point of QCD: STAR capabilities for & \\
\hspace{3em}low $\sqrt{s_{_{NN}}}$ running & T.\ Nayak\\
Low energy operation of RHIC: AGS low energy & \\
\hspace{3em}extraction performance & N.\ Tsoupas\\
Luminosity monitoring issues for low energy RHIC operations& A.\ Drees\\
Electron cooling in RHIC at low energies & A.\ Fedotov\\
Energy dependence of thermal parameters in heavy ion collisions & K.\ Redlich\\
Observable power laws at the QCD critical point & N.\ Antoniou\\
The Compressed Baryonic Matter experiment at FAIR & P.\ Senger\\
Excitation function; experimental perspective & N.\ Xu\\
Experience with CERES & H.\ Appelshauser\\
Critical point at SPS? & R.\ Stock\\
Lattice calculations at finite baryon potential & Z.\ Fodor\\
Fluctuations & V.\ Koch\\
Strangeness and phase changes in relativistic heavy ion collisions & J.\ Rafelski\\
Can we discover the first-order phase transition at RHIC? & J.\ Randrup\\
$v_1$- and $v_2$-flow: Barometry @ HiMu-RHIC - Pinning down & \\
\hspace{3em}the order of the phase transition & H.\ St\"{o}cker\\
Summary/discussion: Prospects for experiments at RHIC & H.-G.\ Ritter  T.\ Roser\\
\br
\end{tabular}
\end{table}

\end{document}